\documentclass[aps,11pt,superscriptaddress,preprintnumbers,
                secnumarabic,nofootinbib]{revtex4}

\usepackage{epsfig}
\usepackage{bbm}

\newcommand{\lbfig}[1]{\refstepcounter{fig} \label{#1} }
\newcounter{fig}

\begin{document} 

\title{\Large Fermion Mass Generation in de Sitter Space}

\author{Bj\"orn~Garbrecht}
\email[]{Bjorn@HEP.Man.Ac.UK}
\affiliation{School of Physics \& Astronomy, University of Manchester,
Oxford Road, Manchester M13 9PL, United Kingdom}

\author{Tomislav Prokopec}
\email[]{T.Prokopec@Phys.UU.NL}
\affiliation{Institute for Theoretical Physics (ITF) \& Spinoza Institute, Utrecht University, Leuvenlaan 4, Postbus 80.195, 3508 TD Utrecht, The Netherlands}


\begin{abstract}
We study the one-loop radiative corrections for massless fermions 
in de Sitter space induced by a Yukawa
coupling to a light, nearly minimally coupled scalar field. We show that
the fermions acquire a mass. Next we construct the corresponding (nonlocal) effective
fermionic action, which -- in contrast to the case of a massive Dirac
fermion -- preserves chirality.
Nevertheless, the resulting fermion dynamics is precisely that of a Dirac fermion with
a mass proportional to the expansion rate.
Our finding supports the view that an observer or a test
particle responds to a scalar field in inflation by shifting its energy
rather than seeing a thermal bath.
\end{abstract}


\maketitle

\section{Introduction}
The question of the dynamics of fermions interacting through a Yukawa
coupling with a massless minimally coupled scalar field in de Sitter space
was originally considered in Ref.~\cite{ProkopecWoodard:2003}.
It is the simplest nontrivial example allowing for a one loop
study of quantum effects during inflation.
Here we reconsider the problem, by assuming that the scalar field is light
and nearly minimally coupled, which allows us to calculate
radiative corrections employing
a de Sitter invariant scalar propagator. We make substantial analytic
progress, such that some of the questions 
that were left unanswered in the original work, {\it e.g.} the problem of Pauli
blocking and vector current conservation, are resolved.

Gaining analytical understanding of fermion dynamics during inflation
may have important cosmological ramifications, in particular for baryogenesis
and the origin of dark matter in the Universe. Closely related are effects in
massless scalar quantum electrodynamics, which may be of relevance for cosmic
magnetogenesis~\cite{Prokopec:PhotonMass,ProkopecPuchwein:2003}, and of graviton
induced self energy~\cite{MiaoWoodard:2005} in de Sitter space.

On a more conceptual level, quantum theory in de Sitter space is most commonly
considered in the context of horizon thermodynamics, which is locally witnessed
by an Unruh detector at a response rate corresponding to
thermal radiation~\cite{GibbonsHawking:1977,GarbrechtProkopec:2004b}.
In recent work, we argued
however that local effects predominantly become manifest in a shift of
the detector's energy levels~\cite{GarbrechtProkopec:2005}.
Since the Unruh detector is an idealised
device, it is of interest what a test particle in a proper
field-theoretic setting experiences, in particular whether effects similar
to those of a thermal bath are present.

\section{One-Loop Self Energy}
 Massless fermions are conformally invariant, such that in a
$D$-dimensional spacetime with the metric
\begin{equation}
ds^2=a^2\left(-d\eta^2+d \mathbf{x}^2 \right)\,,
\end{equation}
their propagator arises from a conformal rescaling of the flat-space
counterpart
\begin{equation}
{\rm i} S (x;x^\prime)=(aa^\prime)^{\frac{1-D}{2}}{\rm i}
{\partial\!\!\!/}\left\{
\frac{\Gamma\left(\frac{D}{2}-1\right)}{4\pi^{\frac{D}{2}}}
\left[\Delta x^2 (x;x^\prime)\right]^{1-\frac{D}{2}}
\right\}
=-\frac{\Gamma\left(\frac{D}{2}\right)}{2\pi^{\frac{D}{2}}}
\frac{(aa^\prime)^{\frac{1-D}{2}}{\rm i}
\gamma^{\mu}\Delta x_\mu}
{\left[\Delta x^2 (x;x^\prime)\right]^\frac{D}{2}}
\,,
\end{equation}
where $\Delta x^2 = -(|\eta-\eta^\prime|-{\rm i}\epsilon)^2 + \|\vec x - \vec x^{\,\prime}\|^2$
and $\Delta x_\mu = x_\mu - x^{\,\prime}_\mu$.

De Sitter space is endowed with the scale factor 
\begin{equation}
  a = -\frac{1}{H\eta}
\,,
\label{scale factor}
\end{equation}
where $H$ denotes the Hubble parameter and $\eta$ conformal time. 
The Green function for a light nearly minimally coupled scalar in de Sitter space is of
the following de Sitter invariant form~\cite{ChernikovTagirov:1968}, 
\begin{equation}
    {\rm i}\Delta(x,x')=\frac{\Gamma(\frac{D-1}{2}+\nu_D)
    \Gamma(\frac{D-1}{2}-\nu_D)}{(4\pi)^{\frac{D}{2}}
    \Gamma(\frac{D}{2})}H^{D-2}\;
{}_2F_1\Big(\frac{D-1}{2}+\nu_D,\frac{D-1}{2}-\nu_D,\frac{D}{2},1-\frac{y}{4}\Big)
\label{Delta:Exact}
\,,
\end{equation}
where
\begin{equation}
\nu_D=\left[\left(\frac{D-1}{2}\right)^2-\frac{m^2+\xi R_D}{H^2}
      \right]^\frac 12
\,,
\label{nuD}
\end{equation}
$m$ is the mass of the scalar field,
$R_D = D(D-1)H^2$ denotes the Ricci scalar curvature of de Sitter space,
and $\xi$ is the coupling constant of the scalar field to
curvature
($\xi=1/6$ corresponds to conformal coupling, $\xi=0$ to minimal coupling).
The function $y=y(x;x^\prime)$ is related to the de Sitter invariant length,
$\ell=\ell(x;x^\prime)$, as follows:
\begin{eqnarray}
 y &=& 4\sin^2\Big(\frac12 H\ell\Big) = aa^\prime H^2 \Delta x^2
\label{y}\,,
\\
\Delta x^2 &=& -(|\eta-\eta^\prime|^2 - i\epsilon)^2 + \|\vec x-\vec x^{\,\prime}\|^2
\label{Delta x2}
\,,
\end{eqnarray}
where $a\equiv a(\eta) = -1/(H\eta)$, $a^\prime \equiv a(\eta^\prime) = -1/(H\eta^\prime)$.

If $m^2 + \xi R_D =0$, the Green function~(\ref{Delta:Exact}) is ill-defined.
When attempting
to calculate it by performing the momentum-sum over the modes, one sees
that this problem corresponds to an infrared
divergence~\cite{TsamisWoodard:1993},
which has to be regulated. There exists no de Sitter invariant regulator~\cite{Allen:1985},
which implies that there is no de Sitter invariant Green function for 
a massless minimally coupled scalar field. 
In a work on scalar electrodynamics in de Sitter space~\cite{ProkopecPuchwein:2003},
the use of a de Sitter invariant propagator with a small mass or weak coupling $\xi$
to gravity was used in studying radiatively induced photon mass generation in de Sitter space.
Here, we adapt this method to study the fermion dynamics in de Sitter
background.

 Since we are primarily interested in considering a small scalar 
mass and weak coupling to gravity,
\begin{equation}
 m^2 + \xi R_D \ll H^2 
,
\label{small mass}
\end{equation}
we find the following representation of the scalar propagator
useful ({\it cf.} also Ref.~\cite{ProkopecPuchwein:2003}):
\begin{eqnarray}
 {\rm i}\Delta(x;x^\prime) &=& \frac{H^{D-2}}{(4\pi)^{\frac{D}2}}
\frac{\Gamma\Big(\frac{D}2\Big)\Gamma\Big(1-\frac{D}2\Big)}
     {\Gamma\Big(\frac{1}2+\nu_D\Big)\Gamma\Big(\frac{1}2-\nu_D\Big)}
    \Bigg\{
          \sum_{n=0}^\infty
           \frac{\Gamma\Big(\frac{D-1}2+\nu_D+n\Big)
                  \Gamma\Big(\frac{D-1}2-\nu_D+n\Big)}
                {\Gamma\Big(\frac{D}2+n\Big)\Gamma\big(n+1\big)}    
           \bigg(\frac{y}{4}\bigg)^{n}
\label{scalar propagator}
\\
       &&\hskip 4.8cm
          -  \sum_{n=-1}^\infty
           \frac{\Gamma\Big(\frac{3}2+\nu_D+n\Big)
                  \Gamma\Big(\frac{3}2-\nu_D+n\Big)}
                {\Gamma\Big(3-\frac{D}2+n\Big)\Gamma\big(n+2\big)}    
           \bigg(\frac{y}{4}\bigg)^{n-\frac{D}2+2}
    \Bigg\}
\,.
\nonumber
\end{eqnarray}
Notice that the conformal propagator corresponds to the $n=-1$ term 
of the second series (containing the $D$-dependent powers of $y$). 
We have recast the scalar propagator into the form~(\ref{scalar propagator}),
such that the cancellation of all terms $n\geq 0$ is manifest in $D=4$.
The cancellation occurs only to leading order in an expansion around 
$D=4$ however. Since the prefactor is singular at $D=4$, the true and finite 
result in $D=4$ is obtained when the series are expanded to linear order
in $D-4$, resulting in coefficients that are suppressed by
$(m^2+\xi R_D)/H^2\ll 1$, as we show below. The explicit form for the 
propagator that is suitable for this problem is thus
($\nu = \big[(3/2)^2-(m^2+\xi R)/H^2\big]^{1/2}$),  
\begin{eqnarray}
 {\rm i}\Delta(x;x^\prime) &=& \frac{H^{D-2}}{4\pi^{D/2}}
                   \Gamma\Big(\frac{D}2-1\Big) \frac{1}{y^{\frac{D}2-1}}
\nonumber\\
&+&\! \frac{H^2}{16\pi^2}
          \sum_{n=0}^\infty
           \frac{\Gamma\Big(\frac32+\nu+n\Big)\Gamma\Big(\frac32-\nu+n\Big)}
                {\Gamma\Big(\frac12+\nu\Big)\Gamma\Big(\frac12-\nu\Big)}    
           \bigg(\frac{y}{4}\bigg)^{n}
    \Big[\ln\Big(\frac{y}4\Big)
       + \psi\Big(\frac32+\nu+n\Big) + \psi\Big(\frac32-\nu+n\Big)
\nonumber\\
&& \hskip 7.2cm
       - \psi\Big(1+n\Big) - \psi\Big(2+n\Big)
    \Big] + {\cal O}(D-4)
\,.\qquad
\label{scalar propagator:D=4}
\end{eqnarray}

The fermionic and scalar propagators can now be assembled to give the
self-energy
\begin{equation}
-{\rm i} \Sigma (x;x')=
\left( -{\rm i} f \mu^{\frac{4-D}{2}} a^D \right) {\rm i} S (x;x')
\left( -{\rm i} f \mu^{\frac{4-D}{2}} a^{\prime D} \right)
{\rm i} \Delta(x;x') +
{\rm i} {\delta Z}_2 (a a')^{\frac{D-1}{2}}
 {\rm i} {\partial\!\!\!/} \delta^D(x-x')
\,,
\end{equation}
where $f$ is a dimensionless Yukawa coupling constant,
$\delta Z_2$ denotes the fermionic field-strength
renormalisation counterterm and $\mu$ is a renormalisation scale.

More explicitly, the one-loop fermion self-energy is given by
\begin{eqnarray}
-{\rm i}\Sigma(x;x^\prime)
\!\!&=&\!\!\frac{f^2 \mu^{4-D}
(a a^\prime)^{\frac32}}{8 \pi^{D}} 
    \Gamma\Bigl(\frac{D}2\Bigr) \Gamma\Bigl(\frac{D}2 - 1\Bigr)
\frac{{\rm i} \gamma^{\mu} {\Delta x}_{\mu}}
     {\Delta x^{2D-2}}
\nonumber\\
&-&\!\!\frac{f^2 H^2 (a a')^{\frac52}}{2^5 \pi^4} \Big(\nu^2 - \frac14\Big)
\frac{{\rm i} \gamma^{\mu} {\Delta x}_{\mu}}{\Delta x^4}
\bigg\{
       \ln\left(\frac{y}4\right)
       + \psi\Big(\frac32+\nu\Big) + \psi\Big(\frac32-\nu\Big)
       + 2\gamma_E - 1
\bigg\}
\nonumber\\
&-&\!\!\frac{f^2 H^2 (a a')^{\frac72}}{2^8 \pi^4} 
            \Big(\nu^2 - \frac14\Big)\Big(\frac94-\nu^2\Big)
\frac{{\rm i} \gamma^{\mu} {\Delta x}_{\mu}}{\Delta x^2}
\bigg\{
       \ln\left(\frac{y}4\right)
       + \psi\Big(\frac52+\nu\Big) + \psi\Big(\frac52-\nu\Big)
       + 2\gamma_E - \frac52
\bigg\}
\nonumber\\
&+& \dots + {\rm i}\delta Z_2 (aa^\prime)^{\frac{D-1}{2}}
{\rm i} {\partial\!\!\!/} \delta^{D}(x-x^\prime)
\label{Sigma:Start}
 \,,
\end{eqnarray}
where $\psi(z) = d[\ln(\Gamma(z))]/dz$, $\gamma_E = 0.57.. \equiv -\psi(1)$
is the Euler constant, $2\gamma_E -1 = - \psi(1) - \psi(2)$,
 $2\gamma_E - 5/2 = - \psi(2) - \psi(3)$, and $\nu\equiv \nu_{D=4}$.
 Note that in~(\ref{Sigma:Start}),
 we kept the first term in general $D$ dimensions, since this term requires 
renormalisation, while the other terms contain only integrable singularities,
and hence in writing them we took the limit $D=4$. The higher order terms we 
neglected in~(\ref{Sigma:Start}) are all suppressed as 
$(m^2+\xi R)/H^2$.

Following Ref.~\cite{ProkopecWoodard:2003}, we use the minimal subtraction
scheme to extract the
contribution from the first term in~(\ref{Sigma:Start}) leading to
an ultraviolet divergence in $D=4$. The counterterm is 
\begin{equation}
\delta Z_2= \frac{f^2}{2^4\pi^{\frac{D}{2}}}
\frac{\Gamma\left(\frac{D}{2}-1\right)}{(D-4)(D-3)}
\,.
\end{equation}
Eventually taking $D \rightarrow 4$, we find the renormalised 
fermion self-energy (up to linear terms in $(m^2+\xi R_D)/H^2)$,
\begin{eqnarray}
\Sigma(x;x^\prime)
\!\!&=&\!\!
- \frac{f^2 (a a^\prime)^{\frac32}}{2^{10} \pi^4}
   {\partial\!\!\!/}\; \partial^4 
\Bigl[ \ln^2(\mu^2 {\Delta x}^2) - 2 \ln(\mu^2 {\Delta x^2}) \Bigr]
\nonumber\\
&&\!
- \frac{f^2 (a a^\prime)^{\frac32}}{2^5 \pi^2}
\ln(a a^\prime) {\rm i} {\partial\!\!\!/} \delta^4(x-x^\prime)
\nonumber \\
&&\!
-\,\frac{f^2 H^2 (a a^\prime)^{\frac52}}{2^9 \pi^4}\Big(\nu^2 - \frac14\Big)
\left\{
{\partial\!\!\!/}\;\partial^2 \ln^2(KH^2 \Delta x^2)
  + 2\ln(aa^\prime){\partial\!\!\!/}\;\partial^2\ln(H^2 \Delta x^2)
\right\}
\nonumber \\
&&\!
+\,\frac{f^2 H^4 (a a^\prime)^{\frac72}}{2^{10} \pi^4}
         \Big(\nu^2 - \frac14\Big)\Big(\frac94 - \nu^2\Big)
\left\{
{\partial\!\!\!/}\, \ln^2(K_1 H^2 \Delta x^2)
  + 2\ln(aa^\prime){\partial\!\!\!/}\,\ln(H^2 \Delta x^2)
\right\}
\,,\qquad
\label{Sigma:Renormalised}
\end{eqnarray}
where we defined 
\begin{eqnarray}
K&=&\frac 14 {\rm exp}\bigg[2\gamma_E-1 
                       + \psi\Big(\frac32+\nu\Big) + \psi\Big(\frac32-\nu\Big)
                      \bigg]\,,
\nonumber\\
K_1&=&\frac 14 {\rm exp}\bigg[2\gamma_E-\frac52 
                       + \psi\Big(\frac52+\nu\Big) + \psi\Big(\frac52-\nu\Big)
                        \bigg]
\,.
\label{K-K1}
\end{eqnarray}
When expanded around $s=(m^2+\xi R)/H^2 = 0$, we find 
\begin{eqnarray}
K&\simeq&\frac 14 {\rm exp}\bigg[\frac12-\Big(\frac32-\nu\Big)^{-1}\bigg]
  \simeq \frac 14 {\rm exp}\bigg[\frac12-\frac{3H^2}{m^2 + \xi R}\bigg]
\,,
\nonumber\\
K_1&\simeq&\frac 14 {\rm exp}\Big[-\frac23\Big]
\,,
\label{K-K1:limit}
\end{eqnarray}
%

Note that the second term in the self-energy~(\ref{Sigma:Renormalised})
breaks de Sitter invariance anomalously. In the flat space
limit, where $H=0$ and the scale factor $a$ is constant, it can always be
removed by either setting $a=1$ or by an appropriate
choice of the renormalisation $\delta Z_2$. In de Sitter space however, the 
momentum scale is continuously shifted, 
such that this term does not vanish at all
times.

\section{Effective Equation of Motion}
In order to study the dynamics of fermions, we shall need the 
retarted self energy,
\begin{equation}
 \Sigma_{\rm ret} = \Sigma^{++} + \Sigma^{+-}
.
\label{Sigma:retarded}
\end{equation}
For the calculation of $\Sigma_{\rm ret}$ the following identities are 
useful (recall an additional {\it minus} sign in $\Sigma^{+-}$),
\begin{eqnarray}
\ln(\alpha \Delta x_{++}^2) - \ln(\alpha \Delta x_{+-})
  \!\!\!&=&\!\!\! 2 {\rm i} \pi \theta(\Delta \eta - r)\theta(\Delta \eta)   
\,,
\nonumber \\
\ln^2(\alpha \Delta x_{++}^2) - \ln^2(\alpha \Delta x_{+-})
  \!\!\!&=&\!\!\! 4 {\rm i} \pi \theta(\Delta \eta - r)\theta(\Delta \eta)   
         \ln|\alpha (\Delta \eta^2 - r^2)|
\label{logs:retarded}
\,,
\end{eqnarray}
where $\theta=\theta(x)$ denotes the Heaviside step function,
$\theta(\Delta \eta^2 - r^2) = \theta(\Delta \eta - r)$, 
 $r \equiv \| \vec x - \vec x^{\,\prime}\|$ and we used
\begin{equation}
 \Delta x_{++}^2 
         = -(|\Delta \eta|- i \epsilon)^2 + \| \vec x - \vec x^{\,\prime}\|^2
\,,\qquad
 \Delta x_{+-}^2 
         = -(\Delta \eta + i \epsilon)^2 + \| \vec x - \vec x^{\,\prime}\|^2
\,.
\label{Delta x++ and +-}
\end{equation}

Using this, we can now write the retarded one-loop fermion self-energy as
\begin{eqnarray}
\Sigma_{\rm ret}(x;x^\prime)
\!\!&=&\!\!
- \frac{f^2 (a a^\prime)^{\frac32}}{2^{8} \pi^3}
   {\rm i}{\partial\!\!\!/}\, \partial^4 
\Bigl\{ \theta(\Delta \eta - r)\theta(\Delta \eta)
              \Big[\ln|\mu^2({\Delta \eta}^2 - r^2)| - 1\Big] 
\Bigr\}
\nonumber\\
&&\!\!\!
- \frac{f^2 (a a^\prime)^{\frac32}}{2^5 \pi^2}
\ln(a a^\prime) {\rm i} {\partial\!\!\!/} \delta^4(x-x^\prime)
\nonumber \\
&&\!\!
-\;\frac{f^2 H^2 (a a^\prime)^{\frac52}}{2^7 \pi^3}\Big(\nu^2 - \frac14\Big)
\Bigl\{
{{\rm i}\partial\!\!\!/}\,\partial^2
            \Big[\theta(\Delta \eta - r)\theta(\Delta \eta)
                       \ln|KH^2  (\Delta \eta^2 - r^2)|\Big]
\nonumber\\
  && \hskip 4cm
        +\, \ln(aa^\prime){\rm i}{\partial\!\!\!/}\;\partial^2
               \Big[\theta(\Delta \eta - r)\theta(\Delta\eta)\Big]
\Bigr\}
\nonumber \\
&&\!\!\!
+\frac{f^2 H^4 (a a^\prime)^{\frac72}}{2^{8} \pi^3}
         \Big(\nu^2 - \frac14\Big)\Big(\frac94 - \nu^2\Big)
\Bigl\{
{\rm i}{\partial\!\!\!/}\, 
               \Big[\theta(\Delta \eta - r)\theta(\Delta\eta)
                    \ln|K_1 H^2 (\Delta \eta^2 - r^2)|
               \Big]
\nonumber\\
  && \hskip 5cm
         +\, \ln(aa^\prime){\rm i}{\partial\!\!\!/}\,
               \Big[\theta(\Delta \eta - r)\theta(\Delta\eta)
               \Big]
\Bigr\}
\label{Sigma:Retarded}
\,.
\end{eqnarray}

 We shall now use the result~(\ref{Sigma:Retarded}) 
to solve the one-loop Dirac equation~\footnote{Note that the sign 
in front of the nonlocal term differs from the sign in 
Ref.~\cite{ProkopecWoodard:2003}. The sign error 
in~\cite{ProkopecWoodard:2003} has been corrected in the corresponding
{\it erratum}.}
\begin{equation}
a^{\frac{3}{2}}{\rm i}{\partial\!\!\!/} a^\frac{3}{2}\psi(x)
-\int d^4 x^\prime \Sigma^{\rm ret}(x;x^\prime)\psi(x^\prime)
=0
\label{Dirac:Modified}
\,.
\end{equation}
Let us define the conformally rescaled wave function 
\begin{equation}
\chi(x) \equiv a^{3/2} \psi(x)
          = {\rm e}^{{\rm i} \vec k\cdot \vec x } \chi(\eta)
\label{chi:more function}
\,,
\end{equation}
which by spatial homogeneity allows for the decomposition 
\begin{equation}
\chi(x^\prime) = {\rm e}^{{\rm i} \vec k\cdot \vec x }
                 {\rm e}^{-{\rm i} \vec k\cdot \Delta\vec x } 
                 \chi(\eta^\prime)
\label{chi:more function:2}
\,,
\end{equation}
such that upon inserting the retarded self-energy~(\ref{Sigma:Retarded}) 
into~(\ref{Dirac:Modified}) and performing the angular integrations, we get  
\begin{eqnarray}
&&{\rm i}{\partial\!\!\!/}\, \chi(x)
+ \frac{f^2}{2^{6} \pi^2}
   {\rm i}{\partial\!\!\!/}\, \partial^4 
 \frac{1}{k}\,
        {\rm e}^{{\rm i} \vec k\cdot \vec x } 
 \int^\eta d\eta^\prime \int_0^{\Delta \eta}dr r\sin(kr) 
              \Big[\ln|\mu^2({\Delta \eta}^2 - r^2)| - 1\Big]\chi(\eta^\prime) 
\nonumber\\
&&\hskip 2cm
+ \frac{f^2 }{2^5 \pi^2}
\Big[\ln(a) {\rm i} {\partial\!\!\!/} \chi(x)
   +  {\rm i} {\partial\!\!\!/}  \Big(\ln(a)\chi(x)\Big)
\Big]
\nonumber \\
&&\!\!
+\;\frac{f^2 H^2 a}{2^5 \pi^2}\Big(\nu^2 - \frac14\Big)
\Biggl\{
{{\rm i}\partial\!\!\!/}\,\partial^2
 \frac{1}{k}\,{\rm e}^{{\rm i} \vec k\cdot \vec x } 
 \int^\eta d\eta^\prime a^\prime\int_0^{\Delta \eta}dr r\sin(kr) 
                      \ln\big[KH^2(\Delta\eta^2 - r^2)\big]\chi(\eta^\prime)
\nonumber\\
  && \hskip 3cm
        +\, \ln(a){\rm i}{\partial\!\!\!/}\;\partial^2
 \frac{1}{k}\,{\rm e}^{{\rm i} \vec k\cdot \vec x } 
 \int^\eta d\eta^\prime a^\prime\int_0^{\Delta \eta}dr r\sin(kr) 
                            \chi(\eta^\prime)
\nonumber\\
  && \hskip 3cm
        +\, {\rm i}{\partial\!\!\!/}\;\partial^2
 \frac{1}{k}\,{\rm e}^{{\rm i} \vec k\cdot \vec x } 
 \int^\eta d\eta^\prime a^\prime\ln(a^\prime)\int_0^{\Delta \eta}dr r\sin(kr) 
                            \chi(\eta^\prime)
\Biggr\}
\nonumber \\
&&\!\!
-\frac{f^2 H^4 a^2}{2^{6} \pi^2}
         \Big(\nu^2 -  \frac14\Big)\Big(\frac94  -  \nu^2\Big)
\Biggl\{
{\rm i}{\partial\!\!\!/}\, \frac{1}{k}\,
        {\rm e}^{{\rm i} \vec k\cdot \vec x } 
 \int^\eta d\eta^\prime  {a^\prime}^2\int_0^{\Delta \eta}dr r\sin(kr) 
                    \ln[K_1 H^2 (\Delta \eta^2 - r^2)]
                         \chi(\eta^\prime) 
\nonumber\\
  && \hskip 4.8cm
         +\, \ln(a){\rm i}{\partial\!\!\!/}\,\frac{1}{k}\,
        {\rm e}^{{\rm i} \vec k\cdot \vec x } 
 \int^\eta d\eta^\prime  {a^\prime}^2\int_0^{\Delta \eta}dr r\sin(kr) 
                         \chi(\eta^\prime) 
\nonumber\\
  && \hskip 4.8cm
         +\, {\rm i}{\partial\!\!\!/}\,
 \int^\eta d\eta^\prime{a^\prime}^2\ln(a^\prime)\int_0^{\Delta \eta}
                 dr r\sin(kr) \chi(\eta^\prime)
\Biggr\}
\label{EOM:1}
\,.
\end{eqnarray}
Upon performing the radial integration this becomes
\begin{eqnarray}
&&{\rm i}{\partial\!\!\!/}\, \chi(x)
      + \frac{f^2}{2^{6} \pi^2}
{\rm i}{\partial\!\!\!/}\, \partial^4\frac{1}{k^3}\,
        {\rm e}^{{\rm i} \vec k\cdot \vec x } 
        {\rm e}^{{\rm i} \vec k\cdot \vec x } 
          \int^\eta\! d\eta^\prime 
          \bigg\{
                 \Big[2\ln(\mu{\Delta \eta}) - 1\Big]
                 \Big[\sin(k\Delta \eta) - k\Delta \eta\cos(k\Delta \eta)\Big]
\nonumber\\
&& \hskip 6.3cm
             + (k\Delta \eta)^2\xi(k\Delta \eta) 
          \bigg\} \chi(\eta^\prime) 
\nonumber\\
&& \hskip 1cm
+ \frac{f^2 }{2^5 \pi^2}
\Big[\ln(a) {\rm i} {\partial\!\!\!/} \chi(x)
   + {\rm i} {\partial\!\!\!/}  \Big(\ln(a)\chi(x)\Big)
\Big]
\nonumber \\
&&\!\!
+\;\frac{f^2 H^2 a}{2^5 \pi^2}\Big(\nu^2 - \frac14\Big)
\Biggl\{
{{\rm i}\partial\!\!\!/}\,\partial^2\frac{1}{k^3}
        {\rm e}^{{\rm i} \vec k\cdot\vec x } 
          \int^\eta d\eta^\prime  a^\prime
          \bigg\{
                 \ln(KH^2\Delta \eta^2)
                 \Big[\sin(k\Delta \eta) - k\Delta \eta\cos(k\Delta \eta)\Big]
\nonumber\\
 &&\hskip 6.9cm             + (k\Delta \eta)^2\xi(k\Delta \eta) 
          \bigg\} \chi(\eta^\prime) 
\nonumber\\
  && \hskip 3cm
        +\, \ln(a){\rm i}{\partial\!\!\!/}\;\partial^2\frac{1}{k^3}
        {\rm e}^{{\rm i} \vec k\cdot \vec x } 
          \int^\eta d\eta^\prime  a^\prime
                  \Big[\sin(k\Delta \eta) - k\Delta \eta\cos(k\Delta \eta)\Big]
                         \chi(\eta^\prime) 
\nonumber\\
 && \hskip 3cm
       +\, {\rm i}{\partial\!\!\!/}\;\partial^2\frac{1}{k^3}
        {\rm e}^{{\rm i} \vec k\cdot \vec x } 
 \int^\eta d\eta^\prime  a^\prime\ln(a^\prime)
                  \Big[\sin(k\Delta \eta) - k\Delta \eta\cos(k\Delta \eta)\Big]
                         \chi(\eta^\prime) 
\Biggr\}
\nonumber \\
&&\!\!\!
-\frac{f^2 H^4 a^2}{2^{6} \pi^2}
         \Big(\nu^2 \!- \! \frac14\Big)\Big(\frac94  \!- \! \nu^2\Big)
\Biggl\{
{\rm i}{\partial\!\!\!/}\, \frac{1}{k^3}
        {\rm e}^{{\rm i} \vec k\cdot \vec x } \! 
 \int^\eta d\eta^\prime  {a^\prime}^2
          \bigg\{
                 \ln(K_1H^2{\Delta \eta}^2)
                 \Big[\sin(k\Delta \eta) - k\Delta \eta\cos(k\Delta \eta)\Big]
\nonumber\\
  && \hskip 7.9cm
             + (k\Delta \eta)^2\xi(k\Delta \eta) 
          \bigg\} \chi(\eta^\prime) 
\nonumber\\
  && \hskip 4.2cm
         +\, \ln(a){\rm i}{\partial\!\!\!/}\,\frac{1}{k^3}
        {\rm e}^{{\rm i} \vec k\cdot \vec x } 
 \int^\eta d\eta^\prime  {a^\prime}^2
             \Big[\sin(k\Delta \eta) - k\Delta \eta\cos(k\Delta \eta)\Big]
                         \chi(\eta^\prime) 
\nonumber\\
  && \hskip 4.1cm
         + {\rm i}{\partial\!\!\!/}\,\frac{1}{k^3}
        {\rm e}^{{\rm i} \vec k\cdot\vec x } \!\!
\int^\eta\!\! d\eta^\prime{a^\prime}^2\ln(a^\prime)
                 \Big[\sin(k\Delta \eta) - k\Delta \eta\cos(k\Delta \eta)\Big]
                         \chi(\eta^\prime) 
\Biggr\} = 0
\,,
\quad\;
\label{EOM:2}
\end{eqnarray}
where we made use of the following integrals,
\begin{eqnarray}
 z^2\xi(z) \!\!\!&=&\!\!\! z^2\int_0^1 dx x \sin(zx)\ln(1-x^2)
\label{xi integral}
\\
  \!\!\!&=&\!\!\! 2\sin(z) - \Big[\cos(z)+z\sin(z)\Big]\Big[{\rm si}(2z)+\frac{\pi}{2}\Big]
         + \Big[\sin(z)-z\cos(z)\Big]\Big[{\rm ci}(2z)\!-\!\gamma_E
                      -\ln\Big(\frac{z}{2}\Big)\Big]
\,,
\nonumber
\end{eqnarray}
where 
\begin{eqnarray}
{\rm si}(z) \!\!\!&=&\!\!\! - \int_z^\infty \frac{\sin(t)}{t}dt 
             = \int_0^z \frac{\sin(t)}{t}dt - \frac{\pi}{2} 
\,,
\nonumber\\
{\rm ci}(z) \!\!\!&=&\!\!\! - \int_z^\infty \frac{\cos(t)}{t}dt 
             = \int_0^z \frac{\cos(t)-1}{t}dt + \gamma_E + \ln(z)
\,.
\label{si ci integrals}
\end{eqnarray}

Because the integrands vanish at the upper limit of integration
at least as $(\Delta \eta)^3 \ln(\Delta\eta)$,  
 the operator $\partial^2 = -(\partial_0^2 + k^2)$
acting on the integrals in~(\ref{EOM:2}) commutes with $\int d\eta^\prime$,
and may be directly applied to the integrands. The result is
\begin{eqnarray}
&&{\rm i}{\partial\!\!\!/}\, \chi(x)
+ \frac{f^2}{2^{5} \pi^2}
   {\rm i}{\partial\!\!\!/}\, (\partial_0^2+k^2) \frac{1}{k}\,
        {\rm e}^{{\rm i} \vec k\cdot \vec x } 
          \int^\eta\! d\eta^\prime 
          \bigg\{
                 2\ln(\mu{\Delta \eta})\sin(k\Delta\eta)
            - \cos(k\Delta \eta)\Big[{\rm si}(2k\Delta \eta)+\frac{\pi}{2}\Big]
\nonumber\\
&& \hskip 6.5cm
 + \sin(k\Delta \eta)\Big[{\rm ci}(2k\Delta \eta)
                          - \gamma_E-\ln\Big(\frac{k\Delta \eta}{2}\Big)\Big]
          \bigg\} \chi(\eta^\prime) 
\nonumber\\
&& \hskip 1cm
+ \frac{f^2 }{2^5 \pi^2}
\Big[\ln(a) {\rm i} {\partial\!\!\!/} \chi(x)
   + {\rm i} {\partial\!\!\!/}  \Big(\ln(a)\chi(x)\Big)
\Big]
\nonumber \\
&&\!\!
-\;\frac{f^2 H^2 a}{2^4 \pi^2}\Big(\nu^2 - \frac14\Big)
\Biggl\{
{{\rm i}\partial\!\!\!/}\,\frac{1}{k}\,
        {\rm e}^{{\rm i} \vec k\cdot\vec x } 
          \int^\eta d\eta^\prime  a^\prime
          \bigg\{
                 \Big[\ln(KH^2\Delta \eta^2)+1\Big]\sin(k\Delta \eta)
\nonumber\\
 &&\hskip 3cm
           - \cos(k\Delta \eta)\Big[{\rm si}(2k\Delta \eta)+\frac{\pi}{2}\Big]
         + \sin(k\Delta \eta)\Big[{\rm ci}(2k\Delta \eta)
                           - \gamma_E-\ln\Big(\frac{k\Delta \eta}{2}\Big)\Big]
          \bigg\} \chi(\eta^\prime) 
\nonumber\\
  && \hskip 3cm
        +\, \ln(a){\rm i}{\partial\!\!\!/}\,\frac{1}{k}\,
        {\rm e}^{{\rm i} \vec k\cdot \vec x } 
          \int^\eta d\eta^\prime  a^\prime
                  \sin(k\Delta \eta)\chi(\eta^\prime) 
\nonumber\\
 && \hskip 3cm
       +\, {\rm i}{\partial\!\!\!/}\,\frac{1}{k}\,
        {\rm e}^{{\rm i} \vec k\cdot \vec x } 
 \int^\eta d\eta^\prime  a^\prime\ln(a^\prime)
                  \sin(k\Delta \eta)\chi(\eta^\prime) 
\Biggr\}
\nonumber \\
&&\!\!\!
-\frac{f^2 H^4 a^2}{2^{6} \pi^2}
         \Big(\nu^2 \!- \! \frac14\Big)\Big(\frac94  \!- \! \nu^2\Big)
\Biggl\{
{\rm i}{\partial\!\!\!/}\, \frac{1}{k^3}
        {\rm e}^{{\rm i} \vec k\cdot \vec x } \! 
 \int^\eta d\eta^\prime  {a^\prime}^2
          \bigg\{
                 \ln(K_1H^2{\Delta \eta}^2)
                 \Big[\sin(k\Delta \eta) - k\Delta \eta\cos(k\Delta \eta)\Big]
\nonumber\\
  && \hskip 7cm
             + (k\Delta \eta)^2\xi(k\Delta \eta) 
          \bigg\} \chi(\eta^\prime) 
\nonumber\\
  && \hskip 4cm
         +\, \ln(a){\rm i}{\partial\!\!\!/}\,\frac{1}{k^3}
        {\rm e}^{{\rm i} \vec k\cdot \vec x } 
 \int^\eta d\eta^\prime  {a^\prime}^2
             \Big[\sin(k\Delta \eta) - k\Delta \eta\cos(k\Delta \eta)\Big]
                         \chi(\eta^\prime) 
\nonumber\\
  && \hskip 3.8cm
         + {\rm i}{\partial\!\!\!/}\,\frac{1}{k^3}
        {\rm e}^{{\rm i} \vec k\cdot\vec x } \!\!
\int^\eta\!\! d\eta^\prime{a^\prime}^2\ln(a^\prime)
                 \Big[\sin(k\Delta \eta) - k\Delta \eta\cos(k\Delta \eta)\Big]
                         \chi(\eta^\prime) 
\Biggr\} = 0
\,,
\qquad
\label{EOM:3}
\end{eqnarray}
where we made use of (writing $z=k\Delta \eta$)
\begin{eqnarray}
 (\partial_0^2 + k^2)z^2 \xi(z)
  = 2 k^2\Big\{
            \!-\! \cos(z)\Big[{\rm si}(2z)+\frac{\pi}{2}\Big]
            \!+ \sin(z)\Big[{\rm ci}(2z) \!-\! \gamma_E
                          \!-\!\ln\Big(\frac{z}{2}\Big)\Big]
            - \frac{d}{dz} \frac{\sin(z)\!-\!z\cos(z)}{z}
         \Big\}
\,,   
\nonumber\\
 (\partial_0^2 + k^2)\Big[2\ln(\alpha z)\Big(\sin(z)-z\cos(z)\Big)\Big] 
      = 2 k^2\Big\{
                   \Big[2\ln(\alpha z)+1\Big]\sin(z)
                + \frac{d}{dz} \frac{\sin(z)-z\cos(z)}{z}
             \Big\}
\label{partial2 acting}
\,.
\quad
\end{eqnarray}
The first term in~(\ref{EOM:3}) is the conformal vacuum contribution
and thus does not grow during inflation. Upon conformal 
rescaling, this part of the self energy can be written as a function of 
$x-x^\prime$, such that it can be analysed 
in momentum space, in which it appears local, as it is standardly done.
The second contribution in~(\ref{EOM:3}) is the conformal anomaly, which 
is local and hence cannot be further simplified. The anomaly grows 
logarithmically with the scale factor, and hence it is subdominant
when compared with the third term, 
which grows as $aa^\prime$ 
and $ aa^\prime\ln(aa^\prime)$. This contribution is the dominant 
one in the limit when $s=(m^2+\xi R)/H^2 \rightarrow 0$, since it contains 
terms that go as $1/s$, while the last (fourth) term 
is of order  $s^0$, and we shall not analyse it further.

At this stage, we note that the fermion apparently 
does not experience the effects
of a de Sitter invariant thermal scalar bath at a temperature $T_H=H/2\pi$.
Since this thermal
bath should also be present for a conformally coupled
scalar~\cite{GibbonsHawking:1977}, its interaction
with the fermion has to be described either by the conformal vacuum or the
conformal anomaly contribution. While the former term does not involve $H$ and
therefore contains no information about de Sitter expansion, also the anomaly
cannot mediate such a phenomenon due to its manifest breaking of de Sitter
invariance.

Keeping -- in accordance with the above discussion --
only the first three terms, Eq.~(\ref{EOM:3}) simplifies to
\begin{eqnarray}
&&({\rm i}\gamma^0\partial_0-\vec{\gamma}\cdot\vec k) \chi(k,\eta)
+ \frac{f^2 }{2^5 \pi^2}
\Big[\ln(a) ({\rm i}\gamma^0\partial_0-\vec{\gamma}\cdot\vec k) \chi(k,\eta)
   + ({\rm i}\gamma^0\partial_0-\vec{\gamma}\cdot\vec k)
                   \Big(\ln(a)\chi(k,\eta)\Big)
\Big]
\nonumber \\
&&\!\!
-\;\frac{f^2 H^2 a}{2^4 \pi^2}\Big(\nu^2 - \frac14\Big)
\Biggl\{
({\rm i}\gamma^0\partial_0-\vec{\gamma}\cdot\vec k)\,\frac{1}{k}\,
          \int^\eta d\eta^\prime  a^\prime
          \bigg\{
                 \Big[\ln(KH^2\Delta \eta^2)+1\Big]\sin(k\Delta \eta)
\nonumber\\
 &&\hskip 3.2cm
           - \cos(k\Delta \eta)\Big[{\rm si}(2k\Delta \eta)+\frac{\pi}{2}\Big]
            + \sin(k\Delta \eta)\Big[{\rm ci}(2k\Delta \eta) 
            - \gamma_E-\ln\Big(\frac{k\Delta \eta}{2}\Big)\Big]
          \bigg\} \chi(\eta^\prime) 
\nonumber\\
  && \hskip 3cm
        +\, \ln(a)({\rm i}\gamma^0\partial_0-\vec{\gamma}\cdot\vec k)
                       \frac{1}{k}\,\int^\eta d\eta^\prime  a^\prime
                  \sin(k\Delta \eta)\chi(\eta^\prime) 
\nonumber\\
 && \hskip 3cm
       +\,({\rm i}\gamma^0\partial_0-\vec{\gamma}\cdot\vec k)
               \frac{1}{k}\int^\eta d\eta^\prime  a^\prime\ln(a^\prime)
                  \sin(k\Delta \eta)\chi(\eta^\prime) 
\Biggr\}
 = 0
\,.
\qquad
\label{EOM:4}
\end{eqnarray}

 Before we proceed, we note that 
\begin{equation}
-({\rm i}\gamma^0\partial_0-\vec{\gamma}\cdot\vec k)
                    \frac{1}{k}\,\theta(\Delta\eta)\sin(k\Delta\eta)
\label{retarded kernel}
\end{equation}
is the retarded Green function of the operator 
      $({\rm i}\gamma^0\partial_0-\vec{\gamma}\cdot\vec k)$,
such that the leading order term (contaning the $\ln(K)$) ${\cal O}(1/s)$ satisfies a local
second order differential equation. The local equation is obtained by 
acting with the operator
$- a({\rm i}\gamma^0\partial_0-\vec{\gamma}\cdot\vec k)/a$
 upon Eq.~(\ref{EOM:4}) (of course, the generic terms of the order $O(s^0)$ 
remain thereby nonlocal), 
\begin{eqnarray}
&&\!\!\!\!\!\Big(a\partial_0\frac{1}{a}\partial_0 + k^2
       +{\rm i}a\Big(\partial_0\frac{1}{a}\Big)\gamma^5 hk\Big) \chi(k,\eta)
 +\, \frac{f^2 }{32 \pi^2}
\bigg[\bigg(a\partial_0\frac{\ln(a)}{a}\partial_0 + \ln(a)k^2
       +{\rm i}a\Big(\partial_0\frac{\ln(a)}{a}\Big)\gamma^5 hk
     \bigg) \chi(k,\eta)
\nonumber \\
&& \hskip 7.3cm
   + \bigg(a\partial_0\frac{1}{a}\partial_0
            + k^2+{\rm i}a\Big(\partial_0\frac{1}{a}\Big)\gamma^5 hk
      \bigg)\Big(\ln(a)\chi(k,\eta)\Big)
\bigg]
\nonumber \\
&& \hskip 1cm
-\,\frac{f^2 H^2 a^2}{16 \pi^2}\Big(\nu^2 - \frac14\Big)
\Biggl\{\Big[2\gamma_E + \psi\Big(\frac{3}{2}+\nu\Big) 
             + \psi\Big(\frac{3}{2}-\nu\Big) \Big]\chi(\eta)
       +\,    [\ln(a)+1]\chi(\eta) 
\nonumber\\
&&\hskip 4.3cm
 +\,\frac{1}{a}\partial_0 \int^\eta d\eta^\prime  a^\prime\ln(H^2\Delta \eta^2)
                                             \chi(\eta^\prime) 
 +\,\frac{2}{a} \int^\eta d\eta^\prime  a^\prime
                   \frac{\cos(k\Delta \eta)-1}{\Delta \eta}
                                             \chi(\eta^\prime) 
\nonumber\\
  && \hskip 4.3cm
        +\, \frac{1}{a}\Big(\partial_0\ln(a)\Big)
                       \int^\eta d\eta^\prime  a^\prime
                  \cos(k\Delta \eta)\chi(\eta^\prime) 
\nonumber\\
  && \hskip 4.3cm
        +\, \frac{1}{a}\Big(\partial_0\ln(a)\Big) ih\gamma^5
                       \int^\eta d\eta^\prime  a^\prime
                  \sin(k\Delta \eta)\chi(\eta^\prime) 
\Biggr\}
 = 0
\,,
\qquad
\label{EOM:5}
\end{eqnarray}
where we assumed that $\chi$ is a helicity eigenvector, 
$\gamma^0 \vec\gamma\cdot\vec k \chi(k,\eta) =  kh\chi(k,\eta)$.

From this form, we infer that the conformal anomaly combines
neatly with the tree level operator to yield 
\begin{equation}
  \Big(1+\frac{f^2}{16\pi^2}\ln(a)\Big)(\partial_0^2 +  k^2)
 -aH\Big(1+\frac{f^2}{16\pi^2}\Big[\ln(a)-\frac32\Big]\Big)\partial_0 
 -{\rm i}aHhk\Big(1+\frac{f^2}{16\pi^2}\Big[\ln(a)-\frac12\Big]\Big)
\,,
\label{anomaly explained}
\end{equation}
such that the conformal anomaly correction to the fermionic wave function
may be neglected when
%
\begin{equation}
 \ln(a) \ll \frac{16\pi^2}{f^2}
\,,
\end{equation}
while  from~(\ref{anomaly explained}), one sees that a
different definition of the scale factor $a$ can always be absorbed in the
field strength renormalisation.
%

The leading order contribution $O(s^{-1})$ to the modified
Dirac equation~(\ref{EOM:5}) 
can be easily extracted from Eq.~(\ref{K-K1:limit}). 
Dropping the anomaly term we get 
\begin{eqnarray}
&&\Big(a\partial_0\frac{1}{a}\partial_0 + k^2
          +{\rm i}a\Big(\partial_0\frac{1}{a}\Big) hk\gamma^5\Big) \chi(k,\eta)
+ a^2\frac{3f^2 H^2}{8 \pi^2}
\bigg[\frac{H^2}{m^2 + \xi R} 
\bigg]\chi(k,\eta)
 = 0
\,.
\qquad
\label{EOM:one over s}
\end{eqnarray}
This is to be compared with the corresponding
expression for a free massive conformally rescaled Dirac-fermion in
an expanding FLRW space, a special case of which is de Sitter inflation,
\begin{equation}
-a({\rm i}{\partial\!\!\!/}-am)\frac 1a ({\rm i}{\partial\!\!\!/}+am)\chi(x)=
\partial_\eta^2 \chi(\eta) + \mathbf{k}^2 \chi(\eta)
-\frac{\partial_\eta a}{a} \partial_\eta \chi(\eta)
-{\rm i} \frac{\partial_\eta a}{a}h|\mathbf{k}|\gamma^5\chi(\eta)
+a^2 m^2 \chi(\eta)
=0
\,.
\label{EOM:massive}
\end{equation}
A comparison of Eqs.~(\ref{EOM:massive}) and~(\ref{EOM:one over s})
suggests that the one-loop radiative corrections due to a Yukawa coupling
to a light scalar field generate an effective mass for the fermion,
\begin{equation}
m_\psi^2 \simeq   \frac{3f^2 H^4}{8 \pi^2(m^2 + \xi R)} 
         \equiv \mu^2H^2
\,.
\label{fermion mass}
\end{equation}
This is the main finding of this paper. A comparison with the
corresponding result for the mass $m_\gamma$
of a photon~\cite{Prokopec:PhotonMass,ProkopecPuchwein:2003}
coupled to a light charged scalar ($e$ denotes electric charge),
\begin{equation}
m_\gamma^2 \simeq   \frac{3e^2 H^4}{4 \pi^2(m^2 + \xi R)}
\,,
\end{equation}
suggests that the dynamical mass generation is a generic feature of de Sitter inflation.
However, as we argue in sections \ref{Inflationary Fermion Dynamics} and \ref{Discussion} below, 
a more subtle analysis is required to fully apprehend all of the ramifications of 
the fermionic mass generation mechanism. 

In the original work Ref.~\cite{ProkopecWoodard:2003},
an instability of the fermionic wave function
was claimed, which is due to an incorrect sign in front of the 
fermion self energy, and hence unphysical. When the 
sign error is corrected~\cite{ProkopecWoodard:2003},
the wave function exibits a behaviour that is similar to the
one reported here and which is consistent
with mass generation in the case of a Yukawa coupling to a massless scalar
field.

\section{Inflationary Fermion Dynamics}
\label{Inflationary Fermion Dynamics}

In this section, we derive analytic solutions to Eq.~(\ref{EOM:one over s})
and discuss their properties as well as some implications.
We first note now that
equations~(\ref{EOM:one over s}--\ref{EOM:massive})
can be reduced a first order system
($\phantom{\;\!}^\prime \equiv d/d\eta$),
\begin{eqnarray}
\label{EOM:massive:2}
 {\rm i}L_h^\prime - hkL_h &=& am_\psi R_h
\,,
\nonumber\\
 {\rm i}R_h^\prime + hkR_h &=& am_\psi L_h
\,,
\label{eom LR}
\end{eqnarray}
where we have decomposed the 4-spinor $\chi$ into a direct product of
chirality and helicity 2-spinors as follows, 
\begin{equation}
 \chi = \left(\begin{array}{c}
                 L_h \cr
                 R_h \cr
             \end{array}\right)
         \otimes \xi_h
\,,
\label{LR spinor}
\end{equation}
and where $\xi_h$ is the helicity 2-eigenspinor,
\begin{equation}
   \hat h \xi_h \equiv \hat {\vec k}\cdot\vec \sigma\; \xi_h
                  = h \xi_h
\,.
\label{helicity eigenspinor}
\end{equation}

It is useful to consider the spinors
\begin{equation}
 u_{\pm h} = \frac{L_h\pm R_h}{\sqrt{2}}
 \,,
\label{u spinor:def}
\end{equation}
sucht that Eqs.~(\ref{EOM:massive:2}) turn into
\begin{eqnarray}
\nonumber
 {\rm i}u_{+h}^\prime - am_\psi u_{+h} &=& hku_{-h}
\,,
\\
 {\rm i}u_{-h}^\prime + am_\psi u_{-h} &=& hk u_{+h}
\label{EOM:massive:3}
\,.
\end{eqnarray}
These can decoupled into two second order equations, 
\begin{eqnarray}
 u_{\pm h}^{\prime\prime} +(k^2 + a^2m_\psi^2) u_{\pm h} 
           \pm i (am_\psi)^\prime u_{\pm h} = 0
\,,
\label{EOM:massive:4}
\end{eqnarray}
which in de Sitter space read
\begin{equation}
 u_{\pm h}^{\prime\prime} 
  + \Bigg(k^2 
      + \frac{\frac 14 - \Big(\frac12\mp{\rm i}\frac{m_\psi}{H}\Big)^2}{\eta^2}
    \Bigg) u_{\pm h} 
  = 0
\,.
\label{EOM:massive:5}
\end{equation}
This is Bessel's equation for the index
\begin{equation}
   \nu_\pm = \frac12 \mp {\rm i} \frac{m_\psi}{H}
\label{EOM:index}
\end{equation}
and has the normalisable solution
\begin{equation}
  u_{+h} = {\rm e}^{i\frac\pi 2 \nu_+}\sqrt{-\frac{\pi k\eta}{4}}
             H_{\nu_+}^{(1)}(-k\eta)
\,.
\label{EOM:massive:solution1}
\end{equation}
Making use of Eq.~(\ref{EOM:massive:3}) and the recursion relation
\begin{equation}
  H^{(i)}_{\nu-1}(z) = \frac{d}{dz}H_\nu ^{(i)}(z)
                     + \frac{\nu}{z}H_\nu ^{(i)}(z)
\qquad(i=1,2)
\,,
\label{recursion:Hankel}
\end{equation}
we obtain the second solution, 
\begin{equation}
  u_{-h} = h{\rm e}^{i\frac\pi 2 \nu_-}\sqrt{-\frac{\pi k\eta}{4}}
             H_{\nu_-}^{(1)}(-k\eta)
\,,
\label{EOM:massive:solution2}
\end{equation}
such that both solution are normalised to \emph{one}, 
\begin{equation}
     \sum_{\pm} |u_{\pm h}|^2 = 1
\,.
\label{Bessel normalisation}
\end{equation}
This normalisation condition follows from the requirement that (in spatially 
homogeneous space-times) the vector current for each helicity sector must be 
conserved (see {\it e.g.} Ref.~\cite{GarbrechtProkopecSchmidt:2002}), 
$\partial_\mu j^\mu_{h\psi} = 0$, a consequence
of Noether's theorem applied to phase transformations of the fermion field.
It therefore holds for any fermion with time-dependent
mass term, and its physical meaning is 
charge conservation. The analogous condition for a scalar field is the
conservation of the Wronskian.

We therefore introduce the phase space charge density associated with the vector current
($j_{h \psi}^0 = \int d^3k f_{0h}/(2\pi)^3$), 
\begin{equation}
 f_{0h} = |L_h|^2 + |R_h|^2 = |u_{+h}|^2 + |u_{-h}|^2 = 1
\,,
\label{normalisation:fermions}
\end{equation}
which is normalised to the vacuum fluctuations contribution.
In order to check explicitly that our solutions satisfy
Eq.~(\ref{Bessel normalisation}),
we shall make use of~(\ref{recursion:Hankel}) and the following 
identities~\cite{GradshteynRyzhik:1965} 
\footnote{The last identity in~(\ref{useful equalities:Hankel}) 
is incorrectly stated in Ref.~\cite{GradshteynRyzhik:1965}.}
\begin{eqnarray}
  \nu_+ + \nu_- &=& 1
\,,\qquad \nu_+^* = \nu_-
\,,
\nonumber\\
  H_{-\nu}^{(i)}(z) &=& {\rm e}^{{\rm i}\pi\nu} H_{\nu}^{(i)}(z)
\,,\qquad (i=1,2)
\nonumber\\
  \big[H_{\nu}^{(1)}(z)\big]^* &=& H_{\nu^*}^{(2)}(z^*)
\,.
\label{useful equalities:Hankel}
\end{eqnarray}
Note first that Eqs.~(\ref{normalisation:fermions}),
(\ref{EOM:massive:solution1}) and~(\ref{EOM:massive:solution2}) imply, 
\begin{eqnarray}
 f_{0h} =  \frac{\pi}{4} \frac{z}{k} 
   \Big[
       {\rm e}^{i(\pi/2)(\nu_{+}-\nu_{-})}
            H^{(1)}_{\nu_+}(z)H^{(2)}_{\nu_-}(z)
     +  {\rm e}^{-i(\pi/2)(\nu_{+}-\nu_{-})}
            H^{(1)}_{\nu_-}(z)H^{(2)}_{\nu_+}(z)
    \Big]
\,,\quad (z \equiv -k\eta > 0)
\,.
\label{f_0h3}
\end{eqnarray}
Making use of Eqs.~(\ref{useful equalities:Hankel}), we then obtain
\begin{eqnarray}
 f_{0h} = -i \frac{\pi z}{4} 
   \Big[
        H^{(1)}_{\nu_- - 1}(z)H^{(2)}_{\nu_-}(z)
     -  H^{(1)}_{\nu_-}(z)H^{(2)}_{\nu_- - 1}(z)
    \Big]
\,.
\label{f_0h4}
\end{eqnarray}
Upon making use of the recursion relation~(\ref{recursion:Hankel}) 
and the Wronskian,
\begin{equation}
  H_{\nu - 1}^{(i)}(z) = \frac{d}{dz}H_{\nu}^{(i)}(z)
                           + \frac{\nu}{z}H_{\nu}^{(i)}(z) 
\qquad (i=1,2)
\,,
\label{Hankel:recursion}
\end{equation}
this immediately reduces to
\begin{equation}
  f_{0h} = 1
\,,
\end{equation}
proving thus Eq.~(\ref{normalisation:fermions}).

Even though there is no
charge generation during de Sitter inflation
by the fermion mass generation mechanism under consideration,
there is particle number generation, which can be calculated 
following Ref.~\cite{GarbrechtProkopecSchmidt:2002} as
\begin{eqnarray}
  n_{\vec k h} &=& \frac{1}{2\omega_k}
               \Big(
                    hk f_{3h}  
                  + a \Re[m_\psi]f_{1h}
                  + a \Im[m_\psi]f_{2h} 
               \Big) + \frac 12 
\label{fermion:number}\,,
\end{eqnarray}
where  $\omega_{k}=\sqrt{k^2 + a^2m_\psi^2}$
and $f_{1h}$, $f_{2h}$ and $f_{3h}$ are the scalar, pseudoscalar and 
pseudovector densities on phase space, respectively, defined as
\begin{eqnarray}
  f_{1h} &=&  -2 \Re\big[L_hR_h^*\big] 
        \;\;  =   - \big(|u_{+h}|^2 - |u_{-h}|^2 \big)
\,,
\label{f1h}
\\
  f_{2h} &=&  2\Im \big[L_{h}^*R_{h}\big]
        \quad\;\,  =   2\Im \big[u_{+h}u_{-h}^*\big]
\,,
\label{f2h}
\\
  f_{3h} &=&  |R_h|^2 - |L_h|^2 
          =   - 2\Re \big[u_{+h}u_{-h}^*\big]
\,.
\label{f3h}
\end{eqnarray}
It is known~\cite{GarbrechtProkopecSchmidt:2002} that the particle number 
density on phase space $n_{\vec kh}$ lies between 0 and 1, in accordance
with Pauli blocking.
Note however, that for adiabatic particle production, which is the case at
hand, one should interpret $n_{\vec kh}$ with care, as it may have very
different properties than a corresponding particle distribution in the
time-independent case. For example, an Unruh detector coupled to a scalar
field in an expanding
background is apparently insensitive to the corresponding scalar expression
for $n_{\vec kh}$~\cite{GarbrechtProkopec:2005,Garbrecht:2005}.
As $n_{\vec kh}$ is derived
by diagonalisation of the Hamiltonian, a quantity with a clear physical meaning
is the energy per mode, $\Omega_{\vec k h}=\omega_{k} n_{\vec kh}$,
which defines the total energy density (the vacuum contribution is subtracted),
$\rho = \sum_{h=\pm 1}\int d^3k\Omega_{\vec k h}/(2\pi)^3$.

For the situation at study, Eq.~(\ref{fermion:number}) reduces during
de Sitter inflation to
\begin{eqnarray}
  n_{\vec k h} &=& \frac 12  
                - \frac{1}{2\sqrt{1+\mu^2/z^2}}
               \Big(
                    2h \Re \big[u_{+h}u_{-h}^*\big]
                  + \frac{\mu}{z}  \big(|u_{+h}|^2 - |u_{-h}|^2 \big)
               \Big) 
\label{fermion:number:2}
\,,
\end{eqnarray}
which we now discuss in the super- and sub-Hubble limits, respectively.
Based on the small argument expansion of the mode functions, 
\begin{eqnarray}
u_{+ h} &\stackrel{k\eta \rightarrow 0-}{\longrightarrow}&
               -\frac{{\rm i}}{\sqrt{2\pi}}
         {\rm e}^{i \frac{\pi}{2}\nu_{+}} \Gamma(\nu_{+})
              \Big(\frac{z}{2}\Big)^{\frac 12 - \nu_{+}}
   \bigg[1 
      + {\rm e}^{-{\rm i} \pi \nu_{+}} \frac{\Gamma(-\nu_{+})}{\Gamma(\nu_{+})}
           \Big(\frac{z}{2}\Big)^{2\nu_{+}}
   \bigg]\Big[1+{\cal O}\big(z^2\big)\Big]
\,,
\nonumber\\
u_{- h} &\stackrel{k\eta \rightarrow 0-}{\longrightarrow}&
              - \frac{{\rm i}h}{\sqrt{2\pi}}
         {\rm e}^{{\rm i} \frac{\pi}{2}\nu_{-}} \Gamma(\nu_{-})
              \Big(\frac{z}{2}\Big)^{\frac 12 - \nu_{-}}
   \bigg[1 
      +   {\rm e}^{-i \pi \nu_{-}} \frac{\Gamma(-\nu_{-})}{\Gamma(\nu_{-})}
           \Big(\frac{z}{2}\Big)^{2\nu_{-}}
   \bigg]\Big[1+{\cal O}\big(z^2\big)\Big]
\,,
\label{sol:inf:late-times2}
\end{eqnarray}
it is easy to obtain the late time limit, $z=-k\eta\rightarrow 0$
of expression~(\ref{fermion:number:2}),
\begin{eqnarray}
  n_{\vec k h} &\stackrel{k\eta \rightarrow 0-}{\longrightarrow}&
                  \frac{1-\tanh(\pi\mu)}2  +{\cal O}(z\ln(z))
\nonumber\\
 &=&\frac{1}{{\rm e}^{2\pi H/m_\psi}+1} +{\cal O}(z\ln(z))
\,,
\label{fermion:number:3}
\end{eqnarray}
which is reminiscent of a Fermi-Dirac distribution of (nonrelativistic) particles with 
(de Sitter) temperature, $T_H = H/2\pi$ and an energy,
$E_\psi\simeq m_\psi$. 
For light particles, $m_\psi\ll H$, we have,  $n_{\vec k h}\simeq 1/2$,
while for heavy particles, $m_\psi\gg H$, the population density is -- as 
expected based on {\it e.g.} adiabatic analysis --
exponentially suppressed, $n_{\vec k h}\simeq {\rm exp}(- 2\pi m_\psi/H)$.
We have thus established a mass generation mechanism during inflation,
which is responsible for particle production on super-Hubble scales,
\begin{equation}
   \frac{k}{a}\ll H
\label{k:superHubble}
\,,
\end{equation}
according to the distribution~(\ref{fermion:number:3}). 

 It would be incorrect to infer from the distribution~(\ref{fermion:number:3}) that 
the population density is thermal, since that distribution applies only for the 
infrared modes, which satisfy Eq.~(\ref{k:superHubble}). 
 To get more information about the spectrum, we need to investigate the momentum
distribution, which can be obtained in the ultraviolet limit $z\gg m_\psi/H$
by an expansion of the particle number~(\ref{fermion:number:2})
in $m_\psi/(Hz)$, which turns out to be
\begin{eqnarray}
n_{\vec k h} &=& \frac{m_\psi^2}{16 z^4H^2}
              + {\cal O}\left(\frac{m_\psi^4}{H^4z^6}\right)
\nonumber\\
             &=& \frac{m_\psi^2 H^2}{(2k/a)^4}
              + {\cal O}\left(\frac{m_\psi^4 H^2}{(k/a)^6}\right)
\,.
\end{eqnarray}
Hence fermions do not acquire an exponentially falling thermal distribution in
the momentum $\vec k$, as one would expect if they were interacting with a
thermal bath of scalar particles.

\begin{figure}[htbp]
\begin{center}
\epsfig{file=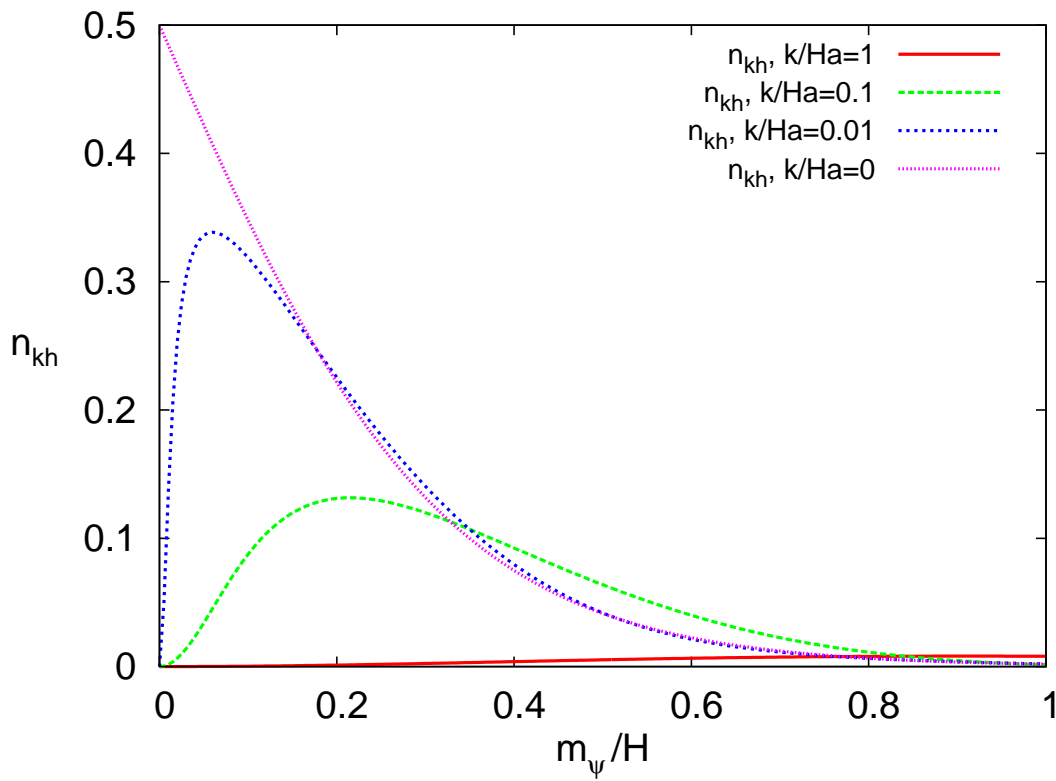, height=4.2in 
       }
\end{center}
\vskip -0.3in
\lbfig{figure 1}
\caption{%
\small
  Fermion particle number $n_{\mathbf{k}h}$ as a function of the fermion mass, $m_\psi/H$.
4 snapshots are shown, $k/(Ha) = 1$ (first Hubble crossing), $k/(Ha) = 0.1, 0.01, 0$
(end of inflation).
}
\vskip -0.in
\end{figure}
 In  FIG.~\ref{figure 1} we show the evolution of the 
particle number, $n_{\vec kh}=n_{\vec kh}(m_\psi/H)$, during de Sitter inflation.
Four snapshots are shown: $k/(Ha) = 1, 0.1, 0.01, 0$. Note that 
most of the particles are created after the first Hubble crossing at $k/a_{1x} = H$. 
At the end of inflation ($k/(Ha)\simeq 0$) the particle number density
approaches the one in Eq.~(\ref{fermion:number:3}).

 An interesting question is what happens to the particles produced during inflation 
in the subsequent epochs of preheating, radiation and matter domination,  
when we expect the fermionic mass to dissolve. This and related questions are 
the subject of a forthcoming publication. 

\section{Discussion}
\label{Discussion}

As main result of this paper, we have presented the effective fermion
mass~(\ref{fermion mass}), dynamically generated in de Sitter space.
From the thermal features of de Sitter space, and especially from the periodicity
of the scalar Green function in imaginary
time~\cite{GibbonsHawking:1977,GarbrechtProkopec:2004}, one might expect that
additional degrees of freedom coupled to the scalar field thermalise. This
would be in analogy with an Unruh detector, which has a thermal response
function and, as a consequence of the principle of detailed balance,
equilibrates at de Sitter temperature. However, we find no indication of
scatterings from a thermal bath or even thermalisation experienced by
the fermion through interaction with the scalar
field. The apparent reason is, that for an Unruh detector, it is assumed that
its internal dynamics is governed by its proper time and 
that detailed balance holds for the detector~\cite{GarbrechtProkopec:2004}.
One might question whether this requirement may in principle be realised by an
experimental
device without disturbing the curved background to an extent which spoils the
measurement. This is a concern which does not apply to the field theoretical
investigation conducted in this work. Nonetheless, the mass generation
mechanism fits into the Unruh detector picture when
relating it to the Lamb shift of the detector's energy
levels~\cite{GarbrechtProkopec:2005},
which can also be interpreted as a radiative correction.

Finally, let us discuss in more detail, how the dynamical mass term is
related to the familiar Dirac fermion masses.
We first note that we can describe the mass generation effect by a
nonlocal effective action, 
\begin{eqnarray}
 S_{\rm eff} = \int d^4 x
               \bigg[a^{3/2} \bar\psi {\rm i}\partial\!\!\!\slash a^{3/2}\psi
                 -  a^{3/2}\bar\psi \frac{a m_\psi^2}{{\rm i}\partial\!\!\!\slash} a^{3/2}\psi
               \bigg] + {\cal O}(s^0)
              + S_{\rm anomaly} 
\,,
\label{effective action:leading order}
\end{eqnarray}
where $m_\psi$ denotes the fermion mass given in Eq.~(\ref{fermion mass}),
$1/({\rm i}\partial\!\!\!\slash)$ is the (retarded) nonlocal operator whose kernel is 
given in~(\ref{retarded kernel}) and 
\begin{equation}
 S_{\rm anomaly} = \int d^4 x
               \bigg[\frac{f^2}{64\pi^2} a^{3/2} \bar\psi 
                    \Big(\ln(a){\rm i}\partial\!\!\!\slash 
                       +  {\rm i}\partial\!\!\!\slash \ln(a)
                    \Big) a^{3/2}\psi
               \bigg]
\label{anomaly action}
\end{equation}
is the effective action for the anomaly.

The question is then, what is the difference between the dynamics induced by
an ordinary fermion mass term, as implied by the standard action, 
\begin{equation}
 S_{\rm massive} =  \int d^4 x
               \Big[a^{3/2} \bar\psi {\rm i}\partial\!\!\!\slash a^{3/2}\psi
                 -  a^4m_\psi\bar\psi\psi
               \Big]
\,,
\label{massive fermion action}
\end{equation}
and the dynamics generated by the nonlocal effective action~(\ref{effective action:leading order}). 
As we have shown above, these two actions lead to identical second order evolution equations
for the chiral densities, $L_h$ and $R_h$ (and likewise for their linear combinations, $u_{\pm h}$). 
Yet there is a difference: the nonlocal action~(\ref{effective action:leading order})
conserves chirality, while its local counterpart violates chirality. 
According to the local action~(\ref{massive fermion action}), the left and right 
handed densities evolve according to the first order system~(\ref{EOM:massive:2}), whose 
graphical representation is shown in terms of mass insertions in FIG.~\ref{figure 2}. 
On the other hand, the nonlocal action~(\ref{effective action:leading order}) cannot 
flip chirality, and its proper graphical representation is shown in FIG.~\ref{figure 3}.
This is the sense in which the fermion mass~(\ref{fermion mass}) is {\it not} 
the genuine Dirac mass, which couples left and right handed amplitudes.
One may think of the equation of motion for the L-handed (R-handed) fermions as being governed
by the system of equations~(\ref{eom LR}), but one should keep in mind that   
the fermions of opposite chirality are ``ghost'' fermions and 
do not comprise physically measurable states. 
This is to be contrasted with the photons in de Sitter background, which 
acquire a mass through interactions with a scalar medium,
and thus an additional longitudinal physical degree of
freedom~\cite{Prokopec:PhotonMass,ProkopecPuchwein:2003}.

\begin{figure}[htbp]
\begin{center}
\epsfig{file=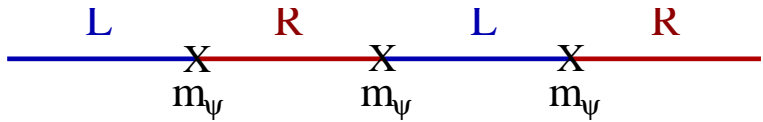, width=4.in
       }
\end{center}
\vskip -0.3in
\lbfig{figure 2}
\caption{%
\small
  Mass insertions for the local fermion Dirac mass term~(\ref{massive fermion action}). 
These insertions violate chirality.
}
\vskip -0.in
\end{figure}
\begin{figure}[htbp]
\begin{center}
\epsfig{file=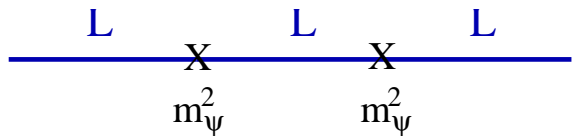, width=3.in
       }
\end{center}
\vskip -0.3in
\lbfig{figure 3}
\caption{%
\small
  Quadratic fermionic mass insertions corresponding to 
the nonlocal effective action~(\ref{effective action:leading order}),
which preserve chirality.
}
\vskip -0.1in
\end{figure}

Putting these findings together suggests dynamical mass generation
for interacting quantum fields as a phenomenon generically
occuring in de Sitter space. Therefore, it is well conceivable
that this mechanism may cause effects during cosmic inflation
which lead to observable signatures.

\end{document}